\documentclass[12pt,preprint]{aastex}
\newcommand{\Lbox}{L_{\mathrm{box}}}
\newcommand{\dgrid}{d_{\mathrm{grid}}}
\newcommand{\rv}{{\bf r}}
\newcommand{\Sigx}{\sigma_{\mathrm{LN}, \, x}}
\newcommand{\SigT}{\sigma_{\mathrm{LN}, \, T}}
\newcommand{\Sign}{\sigma_{\mathrm{LN}, \, n}}
\newcommand{\SigSx}{\sigma_{\mathrm{LN}, \, \mathrm{Sx}}}
\newcommand{\SigSxens}{\sigma_{\mathrm{LN}, \, \mathrm{Sx,ens}}}
\newcommand{\A}{\langle}
\newcommand{\E}{\rangle}

\newcommand{\Ng}{N_{\mathrm{grid}}}

\newcommand{\Dn}{\mbox{$\delta_n$}}
\newcommand{\DT}{\mbox{$\delta_T$}}
\newcommand{\Dx}{\delta_x}

\newcommand{\Sx}{S_{\mathrm{X}}}

\newcommand{\dSx}{\mbox{$\delta_\mathrm{Sx}$}}
\newcommand{\barSx}{\overline{S}_{\mathrm{X}}}
\newcommand{\Sxens}{S_{\mathrm{X,ens}}}
\newcommand{\barSxens}{\A S_{\mathrm{X}} \E }
\newcommand{\dSxens}{\delta_{\mathrm{Sx,ens}}}
\newcommand{\SNt}{\sigma_{\dSxens |_\theta}^2}
\newcommand{\Snt}{\sigma_{\Dnn}^2}

\newcommand{\betafit}{\beta_{\mathrm{X}}}

\newcommand{\rc}{r_{\mathrm{c}}}
\newcommand{\rcfit}{r_{\mathrm{c,X}}}

\newcommand{\lmax}{l_\mathrm{max}}

\newcommand{\thetac}{\theta_\mathrm{c}}

\newcommand{\thetaHPD}{\theta_{\mathrm{HPD}}}
\newcommand{\Rvc}{{\bf R}}

\newcommand{\kv}{{\bf k}}

\newcommand{\thetab}{{\bf \Theta}}

\newcommand{\dAAtt}{d_A^2 \theta^2}
\newcommand{\rceff}{r_{c,\mathrm{eff}}(\theta)}

\newcommand{\Kv}{{\bf K}}
\newcommand{\kmax}{k_{\mathrm{max}}}
\newcommand{\kmin}{k_{\mathrm{min}}}
\newcommand{\kz}{k_l}
\newcommand{\Deff}{\Delta_{\mathrm{eff}}}

\newcommand{\kvd}{{\bf k'}}
\newcommand{\Kvd}{{\bf K'}}
\newcommand{\kmaxd}{k_{\mathrm{max}}'}
\newcommand{\kmind}{k_{\mathrm{min}}'}
\newcommand{\kzd}{k_l'}
\newcommand{\Rvcd}{{\bf R'}}
\newcommand{\alphaq}{\alpha_q}

\newcommand{\alphaDn}{\alpha_{n}}
\newcommand{\alphaDnn}{\alpha_{nn}}
\newcommand{\alphaSx}{\alpha_{\mathrm{Sx}}}
\newcommand{\alphaDT}{\alpha_{T}}
\newcommand{\alphaT}{\alpha_{q,T}}
\newcommand{\Dnn}{\delta_{nn}}
\newcommand{\lprime}{{l^{\prime}}}
\newcommand{\chandra}{{\it Chandra}\ }  
\newcommand{\asca}{{\it ASCA}\ }  

\slugcomment{}
\shortauthors{Kawahara et al.}
\shorttitle{Extracting Cluster Gas Inhomogeneity from X-ray}
\begin{document}
\title{Extracting Galaxy Cluster Gas Inhomogeneity from X-ray Surface
  Brightness: A Statistical Approach and Application to Abell 3667}
\author{
 Hajime Kawahara\altaffilmark{1}, 
 Erik D.~Reese\altaffilmark{1},
 Tetsu Kitayama\altaffilmark{2}, 
 Shin Sasaki\altaffilmark{3}, and
 Yasushi Suto\altaffilmark{1,4}} 
\altaffiltext{1}{Department of Physics, The University of Tokyo, 
Tokyo 113-0033, Japan}
\altaffiltext{2}{Department of Physics, Toho University,  Funabashi,
  Chiba 274-8510, Japan}
\altaffiltext{3}{Department of Physics, Tokyo Metropolitan University,
  Hachioji, Tokyo 192-0397, Japan}
\altaffiltext{4}{Research Center for the Early Universe, Graduate School
of Sciences, The University of Tokyo, Tokyo 113-0033, Japan}
\email{kawahara@utap.phys.s.u-tokyo.ac.jp}
\begin{abstract}

Our previous analysis indicates that small-scale fluctuations in the
intracluster medium (ICM) from cosmological hydrodynamic simulations
follow the lognormal probability density function. In
order to test the lognormal nature of the ICM directly against X-ray
observations of galaxy clusters, we develop a method of extracting
statistical information about the three-dimensional properties of the
fluctuations from the two-dimensional X-ray surface brightness.

We first create a set of synthetic clusters with lognormal
fluctuations around their mean profile given by spherical isothermal
$\beta$ models, later considering polytropic temperature profiles as
well.  Performing mock observations of these synthetic clusters, we
find that the resulting X-ray surface brightness fluctuations also
follow the lognormal distribution fairly well. Systematic analysis of
the synthetic clusters provides an empirical relation between the
three-dimensional density fluctuations and the two-dimensional X-ray
surface brightness.

We analyze \chandra observations of the galaxy cluster Abell 3667, and
find that its X-ray surface brightness fluctuations follow the lognormal
distribution.  While the lognormal model was originally motivated by
cosmological hydrodynamic simulations, this is the first observational
confirmation of the lognormal signature in a real cluster. 

Finally we check the synthetic cluster results against clusters from
cosmological hydrodynamic simulations.  As a result of the complex
structure exhibited by simulated clusters, the empirical relation
between the two- and three-dimensional fluctuation properties
calibrated with synthetic clusters when applied to simulated clusters
shows large scatter.  Nevertheless we are able to reproduce the true
value of the fluctuation amplitude of simulated clusters within a
factor of two from their two-dimensional X-ray surface brightness
alone.

Our current methodology combined with existing observational data is
useful in describing and inferring the statistical properties of the
three dimensional inhomogeneity in galaxy clusters.  

\end{abstract}

\section{Introduction}
\label{sec:intro}

Galaxy clusters have been one of the most important probes of
cosmology (e.g., Bartlett \& Silk 1994; Eke, Cole, \& Frenk 1996; Viana \& Liddle 1996; Kitayama \& Suto 1996, 1997; Kitayama, Sasaki, \& Suto 1998; Holder et al. 2000; Haiman, Mohr, \& Holder 2001; \,\, Majumdar \& Mohr 2004).  In the context
of dark energy surveys, which attracts much of the attention of the
cosmology and particle physics communities, galaxy cluster surveys are
also unique in that they most directly probe the growth of structure
rather than relying solely on distance measurements
\citep[e.g.,][]{albrecht2006}.  In order to capitalize on this, galaxy
cluster surveys, in particular those utilizing the Sunyaev-Zel'dovich
effect \citep[for reviews see, for example,][]{carlstrom02,
birkinshaw99, rephaeli95, sunyaev80}, are currently operating
and many more are planned in the near future.  However, one of the
biggest challenges in interpreting these surveys is relating physical
quantities of galaxy clusters, namely mass, to observable ones.  In
particular, these mass-observable relations may be sensitive to the
inherent complex structure of clusters.  Therefore we must better
understand galaxy clusters to utilize fully the potential of galaxy
cluster surveys in constraining cosmological parameters.

Recent observations of galaxy clusters have revealed a rich variety of
structural complexity.  Recent X-ray satellites with their improved
angular resolution, collecting area, and simultaneous spectral
measurement capabilities have unveiled complex temperature structure
\citep[e.g.,][]{markevitch2000, furusho01}, shock fronts
\citep[e.g.,][]{jones2002}, cold fronts
\citep[e.g.,][]{markevitch2000}, and X-ray holes
\citep[e.g.,][]{fabian2002}.  Improved observational strategies and
analysis methods of lensing observations of galaxy clusters show that
the mass distribution, as opposed to just the gas, is often
complicated as well \citep[e.g.,][]{bradac06}.  Both X-ray and lensing
observations of galaxy clusters reveal that clusters are frequently
undergoing mergers \citep[e.g.,][]{briel04}. With such various and
sundry structural complexities may galaxy clusters reliably be used as
cosmological probes ?

The complex structure seen in galaxy clusters motivates our
investigation of the intracluster medium (ICM) inhomogeneity.  We
note, however, that we take a statistical approach to modeling the
inhomogeneity rather than directly modeling such complex phenomena as
shocks, cold fronts, etc.  Motivated by results from cosmological
hydrodynamic simulations we explore the ramifications of a lognormal
model of the inhomogeneity of the ICM.  This model was first proposed
(Kawahara et al.\ 2007, hereafter Paper I; Kawahara et al.\ 2008), in
this context, to explain the discrepancies between emission weighted
and spectroscopic temperature estimates from galaxy clusters
\citep[][]{mazzotta04, rasia05, vikhlinin06}. They found that local
inhomogeneities of the ICM play an essential role in producing the
systematic bias between spectroscopic and emission weighted
temperatures.

Thus far, the lognormal model has been motivated by and applied only
to clusters from cosmological hydrodynamic simulations. Therefore it
is crucial to see if inhomogeneities in real galaxy clusters also show
the lognormal signature. In reality, this is not a straightforward task
since one can observe clusters in X-rays only through their projection
over the line of sight. Thus we develop a method of extracting
statistical information of the three-dimensional properties of
fluctuations from the two-dimensional X-ray surface brightness.

The rest of the paper is organized as follows. We first summarize the
log-normal model in \S\ref{sec:model}.  We create synthetic clusters
to explore the relationship between the intrinsic cluster
inhomogeneity and X-ray observables in \S\ref{sec:synthetic}.  In \S
\ref{sec:obs} we apply our methodology to \chandra observations of the
galaxy cluster Abell 3667, and then attempt to quantify the nature of
cluster inhomogeneity.  We also compare our synthetic cluster results
with cosmological hydrodynamic simulations in
\S\ref{sec:con}. Finally, we summarize our results in \S\ref{sec:sum}.
Throughout the paper the Hubble constant is parameterized by $h$ in
the usual way, $H_0 = 100\, h$ km s$^{-1}$ Mpc$^{-1}$.

\section{Model of the ICM Inhomogeneity}
\label{sec:model}

\subsection{Lognormal Distribution}
\label{sec:analytic}
 
In order to characterize the inhomogeneity of the ICM, we define the
density and temperature fluctuations as the ratios $\Dn \equiv
n(\rv)/\overline{n}(r)$ and $\DT \equiv T(\rv)/\overline{T}(r)$, where
$n(\rv)$ and $T(\rv)$ are the local density and temperature at radius
$\rv$, and $\overline{n}(r)$ and $\overline{T}(r)$ are the angular average
profiles defined by
\begin{eqnarray}
\label{eq:averagen}
\overline{n}(r) &\equiv& \frac{1}{4 \pi} 
\int n(r,\theta,\phi) \sin\theta\; d\theta\; d\phi\\
\label{eq:averageT}
\overline{T}(r) &\equiv& \frac{1}{4 \pi} \int T(r,\theta,\phi)
\sin\theta\; d\theta\; d\phi
\end{eqnarray}
where $\theta$ and $\phi$ are polar and azimuthal angles,
respectively.   Analysis of hydrodynamical
simulations (Paper I) found that $\Dn$ and $\DT$ are approximately
independent and follow the radially independent lognormal probability
density function (PDF) given by
\begin{equation}
p(\Dx; \Sigx)  \, d \Dx = \frac{1}{\sqrt{2 \pi} \Sigx}
 \exp{\left[ \frac{-\left(\log{\Dx}+\Sigx^2/2 \right)^2}{2 \Sigx^2}
      \right]} \, \frac{d \Dx}{\Dx},
\label{eq:pdf_delta}
\end{equation}
where $x$ denotes $n$ or $T$, $\Dx \equiv x(\rv)/\overline{x}(r)$, and
$\Sigx$ is the standard deviation of the logarithm of density or
temperature.

To construct the two-dimensional surface brightness profile from the
three-dimensional density and temperature distribution, we also need
the properties of the power spectra of the density and temperature
fluctuations.  We adopt statistically isotropic fluctuations with a
power-law type power spectrum for both the density fluctuations
$P_{n}(k) \propto k^{\alphaDn}$ and the temperature fluctuations
$P_{T}(k) \propto k^{\alphaDT}$. These assumptions are based on the
results of the cosmological hydrodynamic simulations described in \S
\ref{subsec:hydro_sim}.

  We use this model to generate synthetic clusters to explore the
relationship between the three-dimensional inhomogeneity in the ICM and
the two-dimensional X-ray surface brightness.

\subsection{Cosmological Hydrodynamic Simulated Clusters}
\label{subsec:hydro_sim}

When one considers the projection of galaxy clusters to two dimensions
for mock X-ray observations, the power spectrum of the fluctuations is
important in addition to the PDF of the inhomogeneity.  Here, we once
again turn to simulations to investigate the power spectrum of the
fluctuations.

We extract the six massive clusters from cosmological hydrodynamic
simulations of the local universe performed by \citet{dolag05}. The
simulations utilize the smoothed particle hydrodynamic (SPH) method, and
assume a flat $\Lambda $ CDM universe with $\Omega_m=0.3, \Omega_b=0.04,
\sigma_8=0.9$, and a dimensionless Hubble parameter $h=0.7$. The number
of dark matter and SPH particles is $\sim 20$ million each within a
high-resolution sphere of radius $\sim 110 $ Mpc, which is embedded in a
periodic box $\sim 343$ Mpc on a side that is filled with nearly seven
million low-resolution dark matter particles. The simulation is designed
to reproduce the matter distribution of the local universe by adopting
the initial conditions based on the {\it IRAS} galaxy distribution
smoothed over a scale of $4.9 h^{-1} \mathrm{Mpc}$. Thus, the six
massive clusters are identified as Coma, Perseus, Virgo, Centaurus,
A3627, and Hydra. A cubic region with 6 $h^{-1}$ Mpc on a side centered
on each simulated cluster is extracted and divided into $512^3$
cells. The density and temperature of each mesh point are calculated
from SPH particles using the B-spline smoothing kernel. A detailed
description of this procedure is given in Paper I. The distance between
two adjacent grid points is given by $ \dgrid = 6 h^{-1}\mathrm{Mpc}/512
\sim 12 h^{-1}$ kpc, which is comparable to the gravitational force
resolution (14 kpc) and the inter-particle separation reached by SPH
particles in the dense centers of clusters. Therefore, the (maximum)
resolution is $\dgrid/\rc \approx 0.1$ assuming $\rc \sim 100$ kpc. This
is about one order of magnitude worse than that of both the synthetic
clusters (\S~\ref{sec:synthetic}) and the observational data
(\S~\ref{sec:obs}).

For each simulated cluster, we compute the radially averaged density
and temperature profiles, $\overline{n}(r)$ and $\overline{T}(r)$,
respectively (Eqn. [\ref{eq:averagen}] and [\ref{eq:averageT}]), and
use them to compute the density and temperature fluctuations $\Dn =
n/\overline{n}$ and $\DT = T/\overline{T}$ at each grid point.  We
extract $128^3$ cells of $\Dn$ and $\DT$ around the center of a
simulated cluster and compute the power spectrum. The distance from the
center to the corner of the $128^3$ cells is $\sim 1.3\ h^{-1}$ Mpc
which is approximately equal to the virial radius of the simulated
clusters ($r_{\mathrm{200}} = 1.0$-$1.6\ h^{-1}$ Mpc). The virial
radius, $r_{\mathrm{200}}$, is the radius within which the mean
interior density is 200 times that of the critical density.

\begin{figure}[!tbh]
  \centerline{\includegraphics[width=85mm]{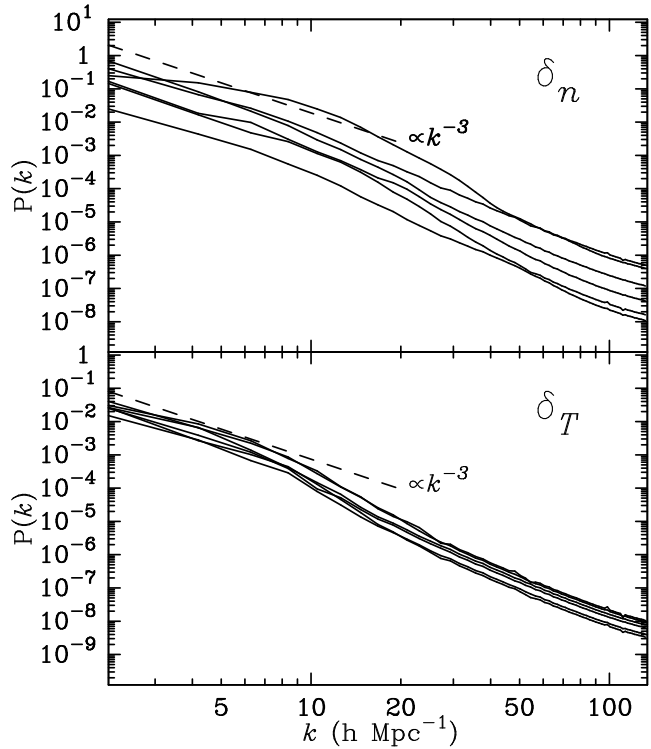}}
  \caption{The power spectra of $\Dn$ (upper) and $\DT$ (lower) of the
    six simulated clusters.  Dashed lines indicate $P(k) \propto
    k^{-3}$. \label{fig:chspower}}
\end{figure}

Figure~\ref{fig:chspower} shows the power spectra for each simulated
cluster for both $\Dn$ (upper panel) and $\DT$ (lower panel).  In each
panel a simple power law, $P(k) \propto k^{-3}$ (dotted line), is also
plotted for comparison.  The power spectra for both the density and
temperature are relatively well approximated by a single power law.
We therefore adopt a power-law spectral model for the density and
temperature fluctuations for the synthetic cluster analysis.

\section{Synthetic Clusters}
\label{sec:synthetic}

 Cosmological hydrodynamic simulations provide a useful test-bed for
exploring cluster structure.  Simulated clusters exhibit complex
density and temperature structure akin to that of real galaxy
clusters.  The resolution of our current simulations, however, is
limited, especially when compared to the resolution available from
current generation X-ray satellites. In addition, we need to
systematically survey the parameter space of $\Sign$ and $\alpha_n$ in
order to relate the X-ray surface brightness fluctuations to the
density fluctuations. Thus we create a set of synthetic clusters at
higher resolution that have lognormal fluctuations around their mean
profile. Analysis of mock observations of these synthetic clusters
enables us to investigate the relation between the X-ray surface
brightness and the statistical properties of the three-dimensional
density fluctuations, namely $\Sign$ and $\alpha_n$.

\subsection{Method \label{ssec:method}}
\subsubsection{Synthetic Cluster Generation}
\label{sss:syn_cl_gen}

The three-dimensional synthetic clusters will be projected to two
dimensions when considering the X-ray surface brightness.  In order to
incorporate a power-law type power spectrum of spatial fluctuations into
the synthetic clusters, we follow a similar methodology as that of
several studies of the interstellar medium \citep{Elmegreen02,FD04}.
First a Gaussian random field with a power-law power spectrum is
constructed and that field is mapped into a lognormal field.  Therefore,
our assumption for the power spectrum is adopted for the Gaussian field
$q$ as opposed to $\delta_n$.  However, we will verify that the ensemble
average of the power spectra of $q$ and $\Dn$ ($ P_q(k) \propto
k^{\alphaq}$ and $ P_{n}(k) \propto k^{\alphaDn}$) have almost the same
power-law indices, $\alphaq \sim \alphaDn$.

We generate the lognormal density fluctuation field as follows.  We
first generate the real random fields, $a({\bf k})$ and
$b({\bf k})$, in $k$-space, whose distribution functions obey
\begin{equation}
\label{eq:deviate1}
{p}(a)da = \frac{1}{ \sqrt{\pi f(k)} }
  \exp{\left[-\frac{a^2}{f (k)}\right]} da,
\quad
{p}(b)db = \frac{1}{ \sqrt{\pi f(k)} }
  \exp{\left[-\frac{b^2}{f (k)}\right]} db, 
\end{equation}
where $f(k) \equiv A k^{\alphaq}$.  Then we compute $q({\bf r})$, the
Fourier transform of a complex field $\tilde{q} ({\bf k}) \equiv a({\bf
k}) + i b({\bf k})$.  With the additional conditions $a({\bf k})=a(-{\bf
k})$ and $b({\bf k})=-b(-{\bf k})$, $q({\bf r})$ becomes a real Gaussian
random field, and its power spectrum, $P_q(k)$, is equal to the input
function $f(k) \equiv A k^{\alphaq}$. The amplitude $A$ is related to
the variance of the Gaussian random field:
\begin{equation}
\sigma_g^2 \equiv  
4 \pi \int_{k_{\rm min}}^{k_{\rm max}} k^2 f(k) dk,
\end{equation}
where $k_{\rm min}$ and $k_{\rm max}$ denote the minimum and maximum
value of the wavenumber.  Finally the lognormal deviate, $\delta_{x}
(\rv)$, is obtained from the Gaussian deviate, $q(\rv)$, using the
relation
\begin{equation}
 \delta_{x} (\rv) =\exp{ \left( \frac{\Sigx}{\sigma_g} q(\rv) -
 \frac{\Sigx^2}{2} \right)},
  \label{eq:lognorm_deviate}
\end{equation}
where $\Sigx$ is the standard deviation of the lognormal field.

We construct synthetic clusters with average density given by the
$\beta$ model and $\Dn$ drawn from a lognormal distribution taking
into account the power-law type power spectrum of spatial
fluctuations.  The $\beta$ model is given by
\citep{cavaliere1976,cavaliere1978}
\begin{equation}
\overline{n}(r) = n_0 \left[ 1 + \left( \frac{r}{\rc} \right)^2
		\right]^{-3 \beta / 2},
  \label{eq:beta_model}
\end{equation}
where $n_0$ is the central electron number density, $\rc$ is the core
radius, and $\beta$ specifies a power-law index.  For simplicity, we first
adopt a fiducial value of $\beta=2/3$, and assume isothermality for
the synthetic clusters.  Later, we examine the effects of varying
$\beta$ (\S~\ref{sss:vary_beta}) and of temperature structure using a
polytropic temperature profile (\S~\ref{ss:tstruct}).

 The density at an arbitrary point is given by
\begin{equation}
n (\rv) = \delta_{n} \overline{n} (r) .
  \label{eq:n_ijk}
\end{equation}
The X-ray surface brightness profile is obtained by projecting the
three-dimensional synthetic cluster down to two dimensions.  For the
isothermal case the projected X-ray surface brightness profile is
\begin{equation}
\Sx (\Rvc) \propto \int [n(\rv)]^2 d l,
  \label{eq:sx_jk}
\end{equation}
where $\Rvc$ indicates the position on the projected plane and $l$ is
the projection of $\rv$ onto the line of sight direction.
\begin{figure}[!tbh]
  \centerline{\includegraphics[width=80mm]{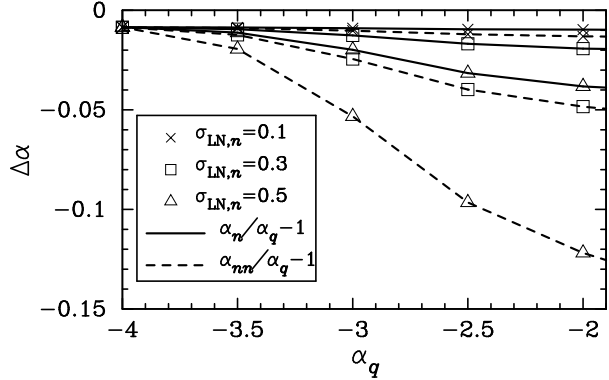}}
\caption{The change of the power-law spectral index of the density
  ($\alphaDn$) and density squared fields ($\alphaDnn$) compared to
  that of the Gaussian field ($\alphaq$). Solid and dashed lines
  indicate $\alphaDn/\alphaq -1$ (density) and $\alphaDnn/\alphaq -1$
  (density squared), respectively.  Each symbol indicates a different
  value of $\Sign$ (cross, square, and triangle correspond to
  $\Sign=0.1$, $0.3$, and $0.5$,respectively.)  The power-law index of
  the density field is very close ($\lesssim 3$\%) to that of the
  Gaussian field used to generate the lognormal distribution and that
  of the square of the density is within $\sim 13$\% for larger values
  of $\Sign$ and $\lesssim 5$\% for smaller values ($\Sign \lesssim
  0.3$). }
  \label{fig:ipchange1}
\end{figure}

Performing the procedure described above, we set up a cubic mesh of
$n(\rv)$ in which our three-dimensional synthetic cluster is located
with $\Ng=512$ grid points along each axis.  We choose the box size
$\Lbox = 10 \, \rc$, which results in the distance between two
adjacent grid points being $\dgrid=10 \, \rc / \Ng \sim 0.02 \, \rc $.

We fit the power spectrum of the $\Dn$ field by a power-law spectrum so
that $P_{n}(\kv) \propto k^{\alphaDn}$. We also fit the power spectrum
of the square density field, $\Dnn \equiv n^2/\A n^2 \E = \Dn^2
\exp{(-\Sign^2)}$ (Appendix B), by the power-law $P_{nn}(k) \propto
k^{\alphaDnn}$, relevant to X-ray surface brightness since $\Sx \propto
\int d\ell \; n^2$.  Throughout this
paper, the notation $\langle x \rangle$ is used to denote the ensemble
average of quantity $x$ over many clusters.

Figure~\ref{fig:ipchange1} shows the change of the power-law spectral
index of the density ($\alphaDn$) and density squared fields
($\alphaDnn$) compared to that of the Gaussian field ($\alphaq$).  The
change in the power-law index for the density and density squared
distributions compared to the initial Gaussian field are small ($<$3\%
and $<$ 13\%, respectively), and therefore, $\alphaq \sim \alphaDn
\sim \alphaDnn$, consistent with the results of \cite{FD04}.

\subsubsection{X-ray Surface Brightness}
\label{ss:em}

To quantify the relationship between the inhomogeneity of the density
and the X-ray surface brightness, $\Sx$, we introduce the X-ray
surface brightness fluctuation from the average radial surface
brightness profile $\barSx (R)$
\begin{equation}
\dSx(\Rvc) \equiv \frac{\Sx(\Rvc)}{\barSx (R)},
  \label{eq:dsx}
\end{equation}
where $R \equiv |\Rvc|$. We define the average
profile $\barSx (R)$ for an individual cluster by fitting the
projected synthetic clusters to an isothermal $\beta$ model
\begin{equation}
\barSx (R) = S_{\mathrm{X},0} \left[ 1 +
  \left(\frac{R}{\rcfit}\right)^2 \right]^{-3\betafit+1/2},
\label{eq:ave1}
\end{equation}
where $S_{\mathrm{X},0}$ is the central X-ray surface brightness,
$\rcfit$ is the core radius, and $\betafit$ specifies the power-law
index for the X-ray surface brightness distribution.  These three
parameters are derived from a model fit to each synthetic cluster. It
is important to emphasize that the average in equation (\ref{eq:ave1})
is defined for {\it an individual cluster}.  We note that if we adopt
directly the average X-ray surface brightness profile instead of a
$\beta$ model fit (Eqn. [\ref{eq:ave1}]), the results are unchanged.
This is because the radial profile is well approximated by the $\beta$
model for the synthetic clusters.  However, for observations of real
galaxy clusters, the $\beta$ model approximation might break down and
one should instead use an average of $\Sx(\Rvc)$ directly in such
cases. In \S 3.2, we will investigate the relation between the
standard deviation of the X-ray surface brightness fluctuations,
$\SigSx$, and that of the intrinsic density fluctuations, $\Sign$.

 Here, we consider the relation of
$\SigSx$ and $\Sign$ for the {\it ensemble average} of clusters
assuming they all obey the $\beta$ model with the same $\beta$, $\rc$,
$\alpha_q$ and $\Sign$:
\begin{eqnarray}
\barSxens (R) &\equiv& \A \Sx (|\Rvc|) \E 
  \label{eq:ens1}\\
\A \Sx (\Rvc) \E &\propto& e^{\Sign^2} \int 
\overline{n}^2 d l, 
  \label{eq:ens2}
\end{eqnarray}
where the exponential term of the right hand side of
equation~(\ref{eq:ens2}) comes from the second moment of the lognormal
distribution (Paper I). Although the ensemble average is {\it not} an
observable quantity, we can describe an analytical prediction of
$\SigSx (R)$ assuming the isothermal $\beta$ model
(Appendix~\ref{a1:den_sb}). In addition, one expects that $\barSx
\sim \barSxens$ if there is a large enough volume compared with the
size of fluctuations when calculating $\barSx$.  In other words, the
spatial average approaches the ensemble average.  For these reasons,
it is useful to consider the ensemble average. Using equations
(\ref{eq:ens1}) and (\ref{eq:ens2}), we define the ensemble average of
fluctuations in the X-ray surface brightness as
\begin{equation}
\dSxens(\Rvc) \equiv \frac{\Sx(\Rvc)}{\barSxens (R)}. 
  \label{eq:ensd}
\end{equation}
We note that the distribution of the square of density fluctuations,
which is proportional to the local emissivity in the isothermal case,
is also distributed according to the lognormal function with a
lognormal standard deviation of $2 \Sign$ if the density fluctuations
follow the lognormal distribution with standard deviation $\Sign$
(Appendix~\ref{sec:a2_densquared}).

\subsection{Statistical Analysis of the Synthetic Clusters}

Here, we investigate the distribution of $\dSx$ of the synthetic
clusters and relate quantities obtainable from observations, $\SigSx$
and $\alphaSx$, to that of the underlying density, $\Sign$ and $\alpha_n$.

\subsubsection{Lognormal nature and the relation between $\SigSx$ and $\Sign$}
\label{ss:synthetic_clusters} 

\begin{figure}[!tbh]
  \centerline{\includegraphics[width=120.0mm]{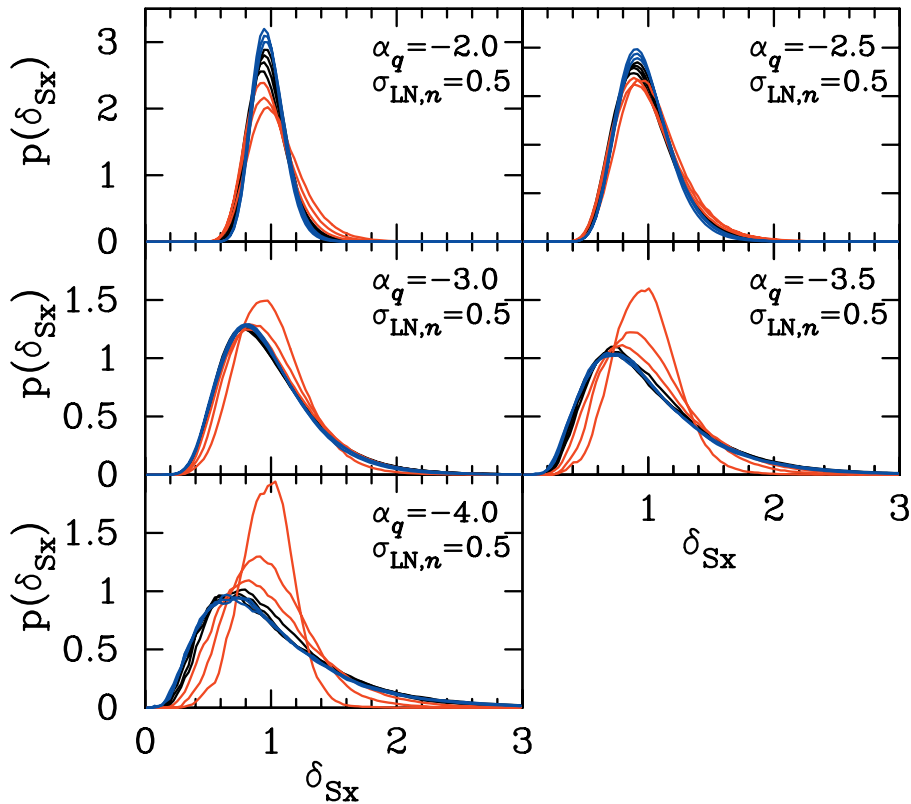}}
\caption{The probability distribution of the ensemble-averaged
distribution of $\dSx$ illustrating the radial dependence.  The
distributions in shells of thickness $0.5 \, \rc$ are shown. Each color
indicates a different radial interval: $R<1.5 \, \rc$ (red), $1.5 \, \rc <
R < 3.5 \, \rc$ (black), and $R>3.5 \, \rc$ (blue).  \label{fig:shells}}
\end{figure}

\begin{figure}[!tbh]
 \centerline{\includegraphics[width=80mm]{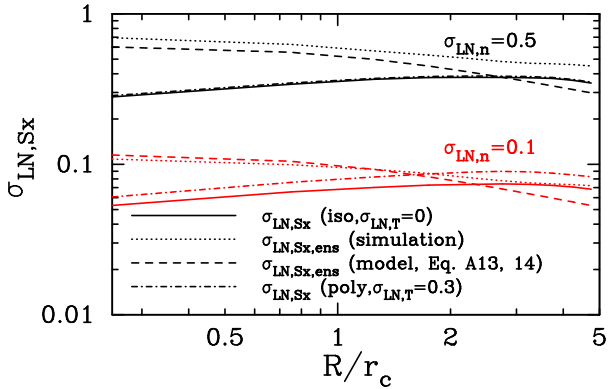}}
\caption{The radial dependence of the standard deviations of the
  logarithm of X-ray surface brightness, $\SigSx$. Two values of
  $\Sign$ are plotted, 0.1 and 0.5, as indicated in the figure. Solid
  and dotted lines show $\SigSx (R)$ calculated using the average
  profile defined by the $\beta$ model (Eq.~[\ref{eq:ave1}]) and the
  ensemble average (Eq.~[\ref{eq:ens2}]), respectively.  Dashed lines
  show the analytical prediction (Eq.~[\ref{eq:thickr}]) . Dash-dotted
  lines indicate the case including the temperature
  structure. Although we show results only for a single power-law
  index, $\alphaq=-3.0$, similar results are obtained in other
  cases. \label{fig:rdt}}
\end{figure}

\begin{figure}[!tbh]
  \centerline{\includegraphics[width=120.0mm]{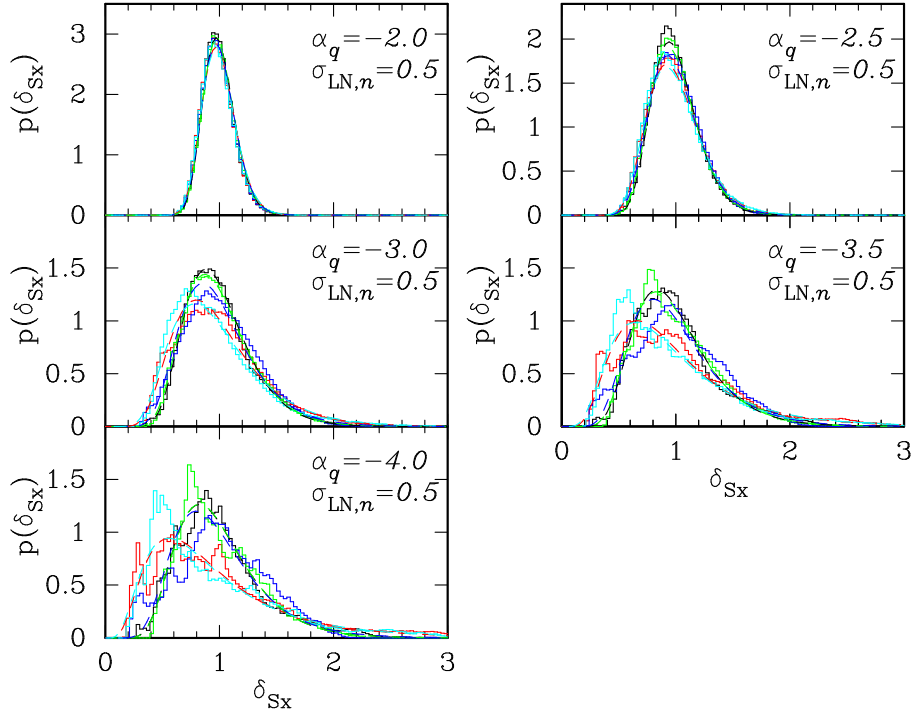}}
 \caption{The probability distribution of $\dSx$ for five individual
 synthetic clusters (solid) along with the best-fit lognormal
 distributions (dashed). Each color shows a different individual
 synthetic cluster.  Each panel shows a different value of the power
 law index of the Gaussian field, $\alphaq$, between $-2$ and $-4$ as
 indicated in each panel.}
 \label{fig:ind}
\end{figure}

We investigate the distribution of $\dSx$ as a function of radial
distance $R$ from the cluster center. We first divide the $\dSx$ field
into shells of thickness $0.5 \, \rc$. The distributions of $\dSx$
within each shell, $p(\dSx ; R)$, averaged over 256 synthetic clusters
are shown in Figure~\ref{fig:shells} for various values of $\alphaq$. We
find that $\dSx$ also approximately follows the lognormal distribution.
The standard deviation of the logarithm of $\dSx$ versus radius, $\SigSx
(R)$, constructed from the averaged shells is displayed in
Figure~\ref{fig:rdt}. Two values of $\Sign$ are plotted, 0.1 and
0.5, in addition to using the average profile defined by both
the $\beta$ model (Eq.~[\ref{eq:ave1}]; solid) and that for the ensemble
(Eq.~[\ref{eq:ens2}]; dotted).  The analytic prediction
(Eq.~[\ref{eq:thickr}]; dashed) and the case including the temperature
structure (\S \ref{ss:tstruct}; dot-dashed) are also plotted. At large
$R$, $\SigSxens (R)$ is approximately $\SigSx (R)$ because the spatial
average tends to the ensemble average due to the large volume used for
averaging.  However, the agreement is poor near the center, where the
ensemble average is not a good approximation.  Although only one value
for $\alphaq$ is shown, similar results are obtained for other values.

Figures ~\ref{fig:shells} and \ref{fig:rdt} indicate that the
probability density function is weakly dependent on the projected
radius $R$.  This radial dependence is caused mainly by two competing
effects.  Consider first the case where the typical nonlinear scale of
fluctuations is much smaller than the size of the cluster itself
(shallow spectrum). As equation (A1) indicates, the surface brightness
at $R$ is given by
\begin{equation}
\Sx(R) \propto \int \Dnn 
\left[1+ \left(\frac{l^2}{\rc^2+R^2}\right)\right]^{-3\beta} dl.
\end{equation}
This implies that the mean value of $\Sx(R)$ is effectively determined
by the integration over the line of sight weighted towards the cluster
center, roughly between $-\sqrt{\rc^2+R^2}$ and $+\sqrt{\rc^2+R^2}$.
This is also true for the variance of $\Sx(R)$. Since the effective
number of independent cells contributing to the variance of $\Sx(R)$
is smaller at smaller projected radii, $\SigSx$ slightly increases for
smaller $R$. This explains the behavior of the shallow spectra results
for $\alpha_q=-2$ and $-2.5$ in Figure~\ref{fig:shells}. On the
contrary, if the typical nonlinear scale of fluctuations is comparable
to or even larger than the cluster size (steep spectrum), the sampling
at the central region significantly underestimates the real
variance. So the $\SigSx$ should increase toward the outer region.
This is seen in Figure 3 for the steeper spectra, $\alpha_q=-3.5$ and
$-4$.

Note the first effect is very small and the second effect becomes
significant only when $\alpha_q < -3$.  The cosmological hydrodynamic
simulations imply that the typical value of $\alpha_q$ is $-3$.
Therefore we neglect the radial dependence of the $\dSx$ field in the
following analysis.
\begin{figure}[!tbh]
 \centerline{
    \includegraphics[width=80mm]{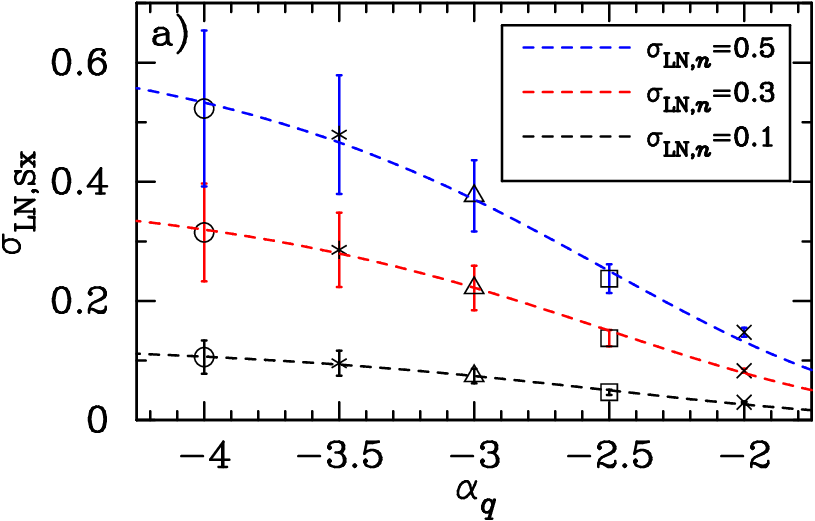}
    \includegraphics[width=83mm]{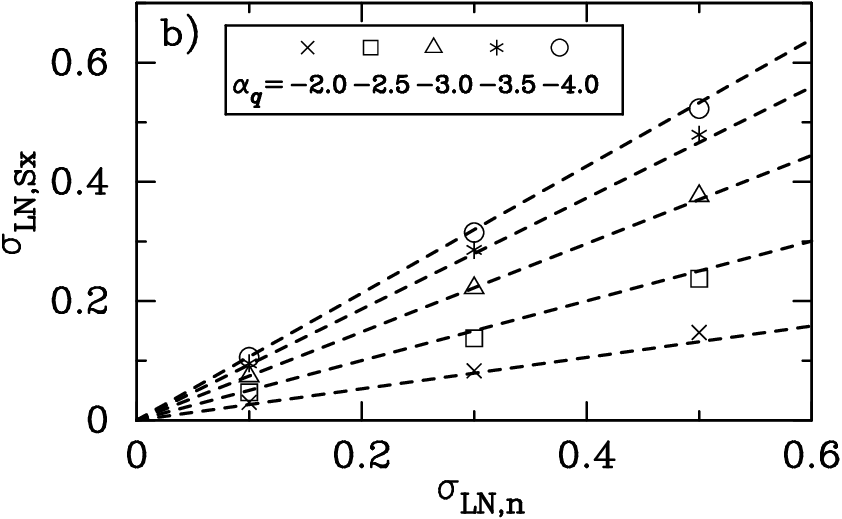}}
\caption{The average of $\SigSx$ from the 256 synthetic cluster sample
as functions of $\alphaq$ (left) and $\Sign$ (right).  The left panel
also shows the standard deviation of $\SigSx$ from the 256 synthetic
clusters and black, red, and blue represent different values of $\Sign$,
$0.1$, $0.3$, and $0.5$, respectively.  In both panels, symbols indicate
values of $\alphaq$ (cross, square, triangle, asterisk, and circle
correspond to $\alphaq= -2$, $-2.5$, $-3$, $-3.5$, and $-4$,
respectively).  Dashed lines show the best-fit approximately linear
$\SigSx$-$\Sign$ relation (Eq.~\ref{eq:fitsKa} and Eq.~ \ref{eq:Kalpha})
for each pair of $\Sign$, $\alphaq$. \vfill \label{fig:mgsigout} }
\end{figure}

From actual observations, we obtain the $\dSx$ map for an individual
cluster, not the ensemble average. Therefore, we evaluate the
distributions of $\dSx$ in individual synthetic clusters.
Figure~\ref{fig:ind} shows the PDF for five individual synthetic
clusters (solid) along with the best-fit lognormal distributions
(dashed). We neglect the radial dependence and use the distribution for
the whole cluster within a diameter of $\Lbox=10 \, \rc$.  Each color
represents a different individual synthetic cluster and each panel shows
a different value of the power-law index of the Gaussian field,
$\alphaq$, with values between -2 and -4.  Even if the analysis is done
for one cluster, the distribution approximately follows the lognormal
distribution.

The noisy behavior for steeper spectra ($\alpha_q=-3.5$, $-4$) in
Figure 5 is due to the presence of fluctuations on scales larger than
that of the cluster, similar to the discussion above for Figure 3.  In
other words, steeper spectra ($\alpha_q<-3$) have relatively more
larger scale fluctuations compared to shallower spectra
($\alpha_q>-3$).  Cosmological hydrodynamic simulations suggest that
$\alpha_q \approx 3$, placing galaxy clusters in the less noisy
regime.  We do not consider the noisy regime further in this paper.

The standard deviations of the logarithm of $\dSx$, $\SigSx$, for the
different sets of $\alphaq$ (symbols) and $\Sign$ (colors) are shown in
Figure~\ref{fig:mgsigout}. The relation between $\Sign$ and $\SigSx$
is approximately linear (right panel) although the proportionality
coefficient depends on $\alphaq$. Therefore, we can write
\begin{eqnarray}
    \label{eq:fitsKa}
    \SigSx = Q(\alphaq) \Sign.
\end{eqnarray}
We find that $Q(\alphaq)$ can be approximated well by the following
function
\begin{equation}
    Q(\alphaq) = \frac{c_1}{c_2 + |\alphaq|^{-4}}.
    \label{eq:Kalpha}
\end{equation}
We calculate the average of $\SigSx/\Sign$ for each $\alphaq$ over
three different values of $\Sign$ ($\Sign=0.1,0.3,$ and $0.5$).  By
fitting $\SigSx/\Sign (\alphaq)$ using equation (\ref{eq:Kalpha}), we
obtain $c_1 = 2.05 \times 10^{-2}$ and $c_2= 1.53 \times 10^{-2}$.

\begin{figure}[!tbh]
 \centerline{\includegraphics[width=65mm]{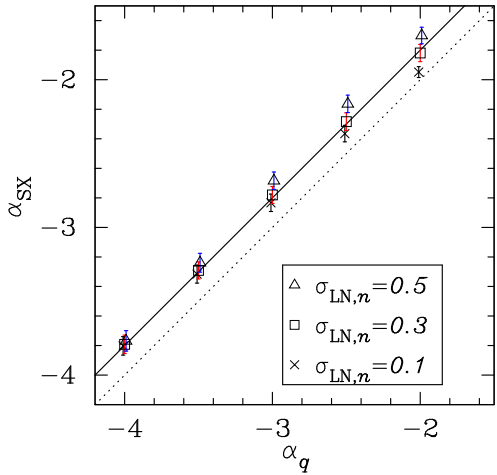}}
\caption{Comparison of the X-ray surface brightness ($\alphaSx$) and
  the input Gaussian field ($\alphaq$) power-law indices.  Symbols and
  error bars indicate the average and the standard deviation,
  respectively, of $\alphaSx$ for 256 samples for different sets of
  $\alphaq$ and $\Sign$.  Symbols correspond to different values of
  $\Sign$, with cross, square, and triangle symbols indicating
  $\Sign=0.1$, $0.3$, and $0.5$, respectively, and the relations
  $\alphaSx=\alphaq$ and $\alphaSx=\alphaq+0.2$ are also shown (dotted
  and solid lines, respectively).  We obtain $\alphaSx$ for each
  individual synthetic cluster by fitting $P_{\Sx}(\Kv)$ of an
  individual cluster under the assumption of both statistical isotropy
  and a power-law ($ \propto K^{\alphaSx}$).
  \label{fig:ipchange2}}
\end{figure}

\begin{figure}[!tbh]
 \centerline{\includegraphics[width=80mm]{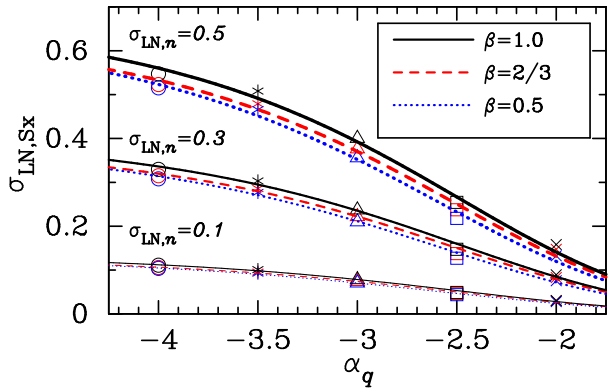}}
\caption{The average of $\SigSx$ over the 256 synthetic clusters as a
  function of $\alphaq$ for different values of the $\beta$ model
  power-law index, $\beta$. Symbols correspond to different values of
  $\alphaq$ as in Figure~\ref{fig:mgsigout}.  Each color shows a
  different value of $\beta$ (black, red, and blue correspond to
  $\beta=1.0, 2/3, $ and $0.5$, respectively). Solid, dashed, and
  dotted lines are fits using equation (\ref{eq:fitsKa}),
  corresponding to $\beta=1.0, 2/3,$ and $0.5$, respectively. The top,
  middle, and bottom sets of three different lines indicate
  $\Sign=0.5, 0.3$, and $0.1$, respectively, as indicated in the
  figure.
\label{fig:betabeta}}
\end{figure}

\subsubsection{Spectral Considerations}
\label{ss:sx_ps}

Because $\SigSx$ is strongly dependent on the power-law index $\alphaq$,
the estimate of $\alphaq$ from the $\dSx$ map is crucial for
interpreting the value of $\SigSx$.  Because $\alphaq$ is an
un-observable quantity, we investigate the relationship between the
power spectra of $\Dn$ and $\dSx$ by fitting the power spectrum of
$\dSx$ under the assumptions of both statistical isotropy and a power
law so that $ P_{\Sx}(\Kv) \propto K^{\alphaSx}$, where $\Kv$ indicates
the two-dimensional wave vector.

Figure~\ref{fig:ipchange2} shows the power-law index of the X-ray
surface brightness, $\alphaSx$, as a function of its counterpart
Gaussian field, $\alphaq$.  Averages and standard deviations over 256
synthetic clusters are shown for three values of the standard
deviation of the logarithm of density, $\Sign$, where crosses,
squares, and triangles correspond to $\Sign$ of $0.1$, $0.3$, and
$0.5$, respectively.  The dotted line corresponds to the relation
$\alphaSx = \alphaq$ and the solid line shows $\alphaSx = \alphaq +
0.2$. We find that $\alphaSx \approx \alphaq + 0.2$ and since $\alphaq
\approx \alphaDn$, this implies $\alphaSx \approx \alphaDn + 0.2$.
This can be understood as follows. As we have seen in
\S~\ref{ssec:method}, the difference between $\alphaDn$ and
$\alphaDnn$ is relatively small ($\lesssim 13$\% and often $\lesssim
5$\%). If one assumes $\dSx$ is the projection of $\Dnn$ (although
this is only strictly true if the average of the surface brightness is
defined by the ensemble average as Eq.[\ref{eq:ens1}]), $\dSx$ can be
described as
\begin{equation}
 \dSx(\thetab) = \int d l \, \Dnn \, W(\thetab,l),
  \label{eq:dsx_theta}
\end{equation}
where $\thetab$ indicates celestial coordinates and $W(\thetab,l)$ is
the window function. If we neglect the $\thetab$-dependence of the
window function and set $W(\thetab,l)=W(l)$, then $P_{\Sx}(\Kv)$ can
be written as
\begin{equation}
P_{\Sx}(\Kv) = \frac{1}{2 \pi} \int  d \kz \, P_{nn} (\kv) \,
|\widetilde{W}(\kz)|^2,
  \label{eq:ps_dsx}
\end{equation}
where $\widetilde{W}(\kz)$ is the Fourier transform of $W(l)$. The
assumption that the size of the cluster is much larger than the typical
scales of the fluctuations yields $|\widetilde{W}(\kz)|^2 \sim 2 \pi
\delta(\kz)$, where $\delta(\kz)$ is the Dirac delta function, and
therefore $K^{\alphaSx} \propto k^{\alphaDnn}$. Thus, we find $\alphaDn
\sim \alphaDnn \sim \alphaSx $ ($\sim \alphaq$).

In this section, we have found that, in principle, one can estimate the
value of $\Sign$ from analysis of X-ray observations.  From the
observations one measures $\SigSx$ and $\alphaSx$ and uses them to
infer $\Sign$, noting that $\alphaq = \alphaSx - 0.2$.  Therefore, one
can estimate the statistical nature of the intrinsic three dimensional
fluctuations from two dimensional X-ray observations.

\subsection{Potential Systematics}
\label{ss:pot_sys}

Using mock observations of isothermal $\beta$ models we found a
relation between the intrinsic inhomogeneity of the three dimensional
cluster gas and the fluctuations in the X-ray surface brightness.  We
turn our attention to the effects of departures from this idealized
model.

\subsubsection{$\beta$ Model Power-law Index}
\label{sss:vary_beta}

In the above description, we have fiducially assumed the $\beta$ model
power-law index $\beta=2/3$. We investigate two other cases,
$\beta=0.5$, and $\beta=1.0$, in Figure~\ref{fig:betabeta}, where we
show $\SigSx$ as a function of $\alphaq$ for different cases of $\beta$
(colors).  The corresponding fits using equation (\ref{eq:fitsKa}) are
also shown.  Although $\SigSx$ tends to increase with increasing
$\beta$, the change is relatively small ($<10$\%).

\subsubsection{Temperature Structure}
\label{ss:tstruct}

\begin{figure}[!tbh]
 \centerline{\includegraphics[width=70.0mm]{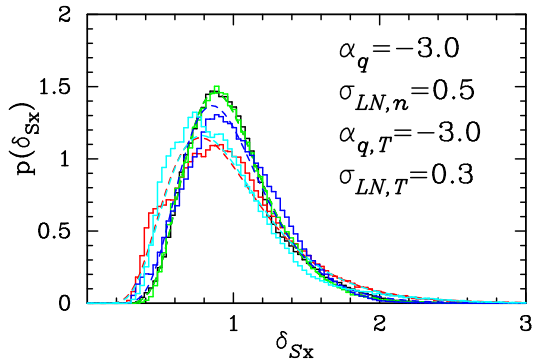}}
 \caption{The distribution of $\dSx$ for five individual clusters
 including the effects of temperature structure. Synthetic clusters
 (solid histogram) and best-fit lognormal model (dashed lines) are
 both shown for each cluster.  Each color corresponds to a different
 individual synthetic cluster.  Although we display only one example
 of the power-law index, $\alphaq=-3.0$, similar results are also
 obtained in other cases. }
\label{fig:tind}
\end{figure}

In the above discussion, we assumed isothermality for the ICM.
However, the X-ray surface brightness also depends on the underlying
cluster temperature structure, including a non-isothermal average
temperature profile and local inhomogeneity. We investigate these
effects for the X-ray surface brightness distribution.

We assume a polytropic profile for the temperature radial
distribution expressed as
\begin{equation}
\overline{T}(r) = T_0 \left(\frac{\overline{n}(r)}{n_0}\right)^{\gamma-1},
  \label{eq:t_polytrope}
\end{equation}
with polytropic index $\gamma = 1.2$ and $T_0 = 6$ keV, which is the
typical set of values in simulated clusters (Paper I). The ensemble
average of the power spectrum of $\DT$ is assumed to have a power-law
form ($\A P_{T}(k) \E \sim P_q(k) \propto k^{\alphaT}$). Because
$\alphaDT \approx \alphaT$ for the same reasons as described in
\S~\ref{ssec:method} for density fluctuations, we fiducially adopt the
power-law index $\alphaT=-3$ based on the results of cosmological
hydrodynamic simulations (for details see \S~\ref{subsec:hydro_sim}).

\begin{figure}[!tbh]
 \centerline{\includegraphics[width=80mm]{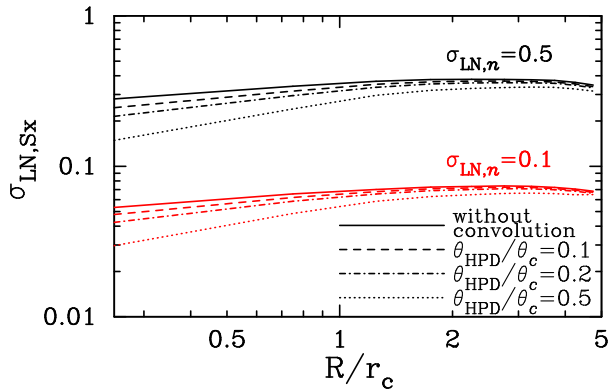}}
\caption{The effect of the PSF on $\SigSx$ as a function of radius,
$R/\rc$, for the case of $\alphaq=-3$ and $\beta=2/3$. Solid curves show
$\SigSx(R)$ without convolution of the PSF. Dashed, dash-dotted, and dotted
curves correspond to $\thetaHPD/\thetac=0.1,0.2$ and $0.5$,
respectively. Two values of $\Sign$ are plotted, 0.1 and 0.5, as
indicated in the figure. 
} \label{fig:psf}
\end{figure}

We create the lognormal distribution $\delta_{T}$ for temperature
fluctuations in the same manner as for the density fluctuations
described in \S~\ref{ssec:method}.  The temperature of an arbitrary
point is assigned according to
\begin{equation}
T(\rv) = \delta_{T}(\rv) \overline{T} (r).
  \label{eq:t_ijk}
\end{equation}
We adopt $\SigT=0.3$, because it is the typical value for simulated
clusters (Paper I).  In addition, we assume that $\Dn$ and $\DT$ are
distributed independently, following Paper I.  The X-ray surface
brightness is given by
\begin{equation}
\Sx (\Rvc) \; \propto  \int  [n(\rv)]^2 \, \Lambda[T(\rv)] \, dl,
  \label{eq:sx_jk2}
\end{equation}
where $\Lambda(T)$ is the X-ray cooling function.  We calculate
$\Lambda(T)$ in the energy range 0.5-10.0 keV using SPEX 2.0
\citep{1996uxsa.conf..411K} on the assumption of collisional
ionization equilibrium and a constant metallicity of 30\% solar
abundances.

Examples of the distribution of $\dSx$ in individual clusters are shown
in Figure~\ref{fig:tind} (solid histogram) along with the best fit
lognormal distributions (dashed lines).  Each color corresponds to a
different individual synthetic cluster.  Although only one value for the
power-law index, $\alpha_q=-3$, is shown, similar results are obtained for
other values.  The radial dependence of $\SigSx$ including the effects
of temperature structure is shown in Figure~\ref{fig:rdt} (dot-dashed).

There are only small differences between the isothermal and
non-isothermal cases. The X-ray surface brightness depends on the
density squared but roughly as $\sqrt{T}$ for bremsstrahlung emission.
Therefore, the temperature structure effects on $\dSx$ are much less
important than those of the density structure.  Hereafter, we neglect
the effects of temperature structure and focus only on the effects of
density inhomogeneity.

\subsubsection{Finite Spatial Resolution}
\label{ss:resolution}

Actual observations by X-ray satellites have finite spatial
resolution, characterized by the point spread function (PSF). We
assume that the PSF is a circularly symmetric Gaussian with standard
deviation $\sigma$.  The PSF can then be parameterized by a single
parameter called the {\it half power diameter} ($\thetaHPD$) in which
50\% of the X-rays are enclosed ($\thetaHPD/\sigma = 2 \sqrt{2 \log
2}$).
 We investigate three cases, $\thetaHPD/\thetac=0.1,0.2$ and $0.5$.
Figure~\ref{fig:psf} shows the effect of the PSF on $\SigSx$ as a
function of radius.  In each case, the average over 256 synthetic
clusters is shown.  Results for no PSF correction ($\thetaHPD=0$, solid)
and $\thetaHPD/\thetac=0.1$ (dashed), 0.2 (dot-dashed), and 0.5 (dotted)
are shown.  As $\thetaHPD/\thetac$ increases, $\SigSx$ near the center
of the cluster decreases.  This can be understood as follows.  In each
radial shell, fluctuations smaller than roughly the radius of the shell
predominately contribute to the fluctuations, namely $\SigSx(R)$.  The
PSF effectively smooths out the smaller scale fluctuations (roughly up
to the size of the PSF), reducing $\SigSx$, while preserving the large
scale fluctuations.  Since the inner shells only contain small scale
fluctuations, they are more strongly affected by the PSF.  The case of
$\thetaHPD/\thetac=0.5$ best illustrates these effects.  The reduction
of $\SigSx$ from the PSF is seen at all radii.  However, it is only a
slight reduction at large radii, increasing as the radius decreases,
with a very large effect near the cluster center.

In summary, when $\Dn$ in three dimensions follows the lognormal
distribution, $\dSx$ in two dimensions also approximately follows the
lognormal distribution.  The mean value of $\SigSx$ for an individual
cluster is strongly dependent on both $\Sign$ and $\alphaq$. Because
$\alphaq$ is approximately equal to $\alphaSx$, in principle, one can
infer $\Sign$ from $\SigSx$ although there is still some dispersion even
if $\alphaq$ is known. In addition, the effect of the temperature
structure is minimal.

\section{Application to Abell 3667}
\label{sec:obs}

Simulations suggest that the lognormal model (Eq.~[\ref{eq:pdf_delta}])
is a reasonable approximation of the small scale structure in galaxy
clusters.  We compare this model with \chandra X-ray observations of the
nearby galaxy cluster Abell 3667 at a redshift $z=0.056$
\citep{struble1999}.  A3667 is a well observed nearby bright galaxy
cluster that does not exhibit a cool core observed by \chandra.  With
its complex structure, including a cold front \citep{vikhlinin2001} and
possible merger scenario \citep[e.g.,][]{knopp96}, A3667 will serve as a
difficult test case for the lognormal model of density fluctuations.

\begin{figure}[!tbh]
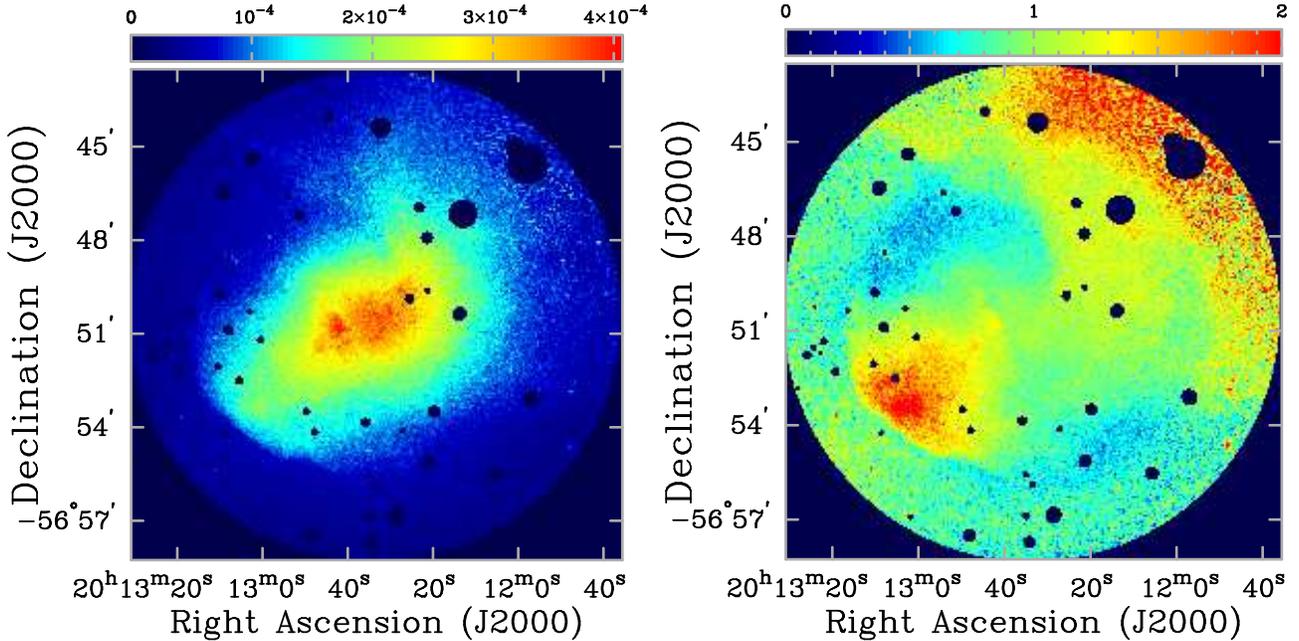

  \centerline{
    \includegraphics[height=8.5cm]{f11a.ps}
    \includegraphics[height=8.5cm]{f11b.ps}
  }
  \caption{\chandra image of the galaxy cluster Abell 3667 (left)
  and the corresponding $\dSx$ image (right). The
  counts image has been divided by the exposure map to yield X-ray
  surface brightness (cnt s$^{-1}$ cm$^{-2}$ arcmin$^{-2}$), including
  scaling for the pixel size.  Point sources in the field have been masked.
}
  \label{fig:a3667_image}
\end{figure}

\begin{deluxetable}{cccc}
  \tablewidth{0pt}
  \tablecolumns{4}
  \tablecaption{A3667 \chandra Observations
  \label{tab:a3667_data}}
  \tablehead{
    \colhead{}      & \colhead{$t_{exp}$} & \colhead{RA} & \colhead{DEC} \\
    \colhead{obsID} & \colhead{(ks)} & \colhead{(h m s)} & \colhead{(d m s)}
  }
  \startdata
  $\phn513$ & $\phn45$ & $20\ 12\ 50.30$ & $-56\ 50\ 56.99$\\
  $\phn889$ & $\phn51$ & $20\ 11\ 50.00$ & $-56\ 45\ 34.00$\\
  $5751$    & $131$    & $20\ 13\ 07.25$ & $-56\ 53\ 24.00$\\
  $5752$    & $\phn61$ & $20\ 13\ 07.25$ & $-56\ 53\ 24.00$\\
  $5753$    & $105$    & $20\ 13\ 07.25$ & $-56\ 53\ 24.00$\\
  $6292$    & $\phn47$ & $20\ 13\ 07.25$ & $-56\ 53\ 24.00$\\
  $6295$    & $\phn50$ & $20\ 13\ 07.25$ & $-56\ 53\ 24.00$\\
  $6296$    & $\phn50$ & $20\ 13\ 07.25$ & $-56\ 53\ 24.00$
  \enddata
\end{deluxetable}

\subsection{Data Reduction}
\label{subsec:obs_data_reduce}

\chandra observations of the galaxy cluster A3667 are summarized in
Table~\ref{tab:a3667_data}.  Listed are the observation identification
numbers, exposure times, and pointing centers of each of the eight
archival \chandra observations of A3667 used in this analysis.  The data
are reduced with CIAO version 4.0 and calibration data base version
3.4.2.  The data are processed starting with the level 1 events data,
removing cosmic ray afterglows, correcting for charge transfer
inefficiency and optical blocking filter contamination, and other
standard corrections, in addition to generating a customized bad pixel
file.  The data are filtered for \asca grades 0, 2, 3, 4, 6 and status=0
events and the good time interval data provided with the observations
are applied.  Periods of high background count rate are excised using an
iterative procedure involving creating light curves in background
regions with 500 s bins, and excising time intervals that are in excess
of 4 $\sigma$ from the median background count rate.  This sigma
clipping procedure is iterated until all remaining data lie within 4
$\sigma$ of the median.  The final events list is limited to energies
0.7-7.0 keV to exclude the low and high energy data that are more
strongly affected by calibration uncertainties.  Finally, the images are
binned by a factor of eight, resulting in a pixel size of 3.94\arcsec.
This pixel size matches the resolution of the synthetic clusters
considered in \S\ref{sec:synthetic}.  In particular, the ratio of pixel
size to the cluster core radius of the \chandra image is similar to the
synthetic cluster grid spacing compared to the synthetic cluster core
radius, namely, for $\theta_\mathrm{c} \sim 180\arcsec$
\citep{rb02,knopp96}, $\theta_{\mathrm{pix}} / \theta_\mathrm{c} \sim
\dgrid / \rc \sim 0.02$.  Exposure maps are constructed for each
observation at an energy of 1 keV. The binned images and exposure maps
for each observation are then combined to make the single image and
exposure map used for the analysis.

A wavelet based source detector is used to find and generate a list of
potential point sources.  The list is examined by eye, removing bogus
or suspect detections, and then used as the basis for our point source
mask.  Figure~\ref{fig:a3667_image} (left) shows the \chandra merged
image of A3667, the counts image divided by the exposure map, where
the point source mask has been applied.  Also shown is the $\dSx$
image (right), discussed below.  A cold front \citep{vikhlinin2001} is
clearly visible in the south-eastern region of the $\dSx$ image.

\subsection{Analysis and Results}
\label{subsec:obs_analysis}

In order to determine the center of A3667, a $\beta$ model is fit to the
data with fixed core radius ($180\arcsec$) and $\beta$ (2/3), using
software originally developed for the combined analysis of X-ray and
Sunyaev-Zel'dovich effect observations
\citep{reese00,reese02,bonamente06}.  Because A3667 is nearby and
appears very large, \chandra observations do not encompass the entire
cluster but provide a wealth of information on the complexities inherent
in galaxy cluster gas.  By using a $\beta$ model fit to the diffuse
emission of the cluster gas we obtain a better measurement of its center
than simply using the brightest pixel or other simple estimates, which
fail to take into account the complex structure manifest in this
cluster.  A circular region of radius $\sim 8\arcmin$ centered on A3667
is used in the analysis, corresponding to two and a half times the
cluster's core radius, the largest usable region from the arrangement of
the combined \chandra observations.

\begin{figure}[!tbh]
  \centerline{\includegraphics[height=6.3cm]{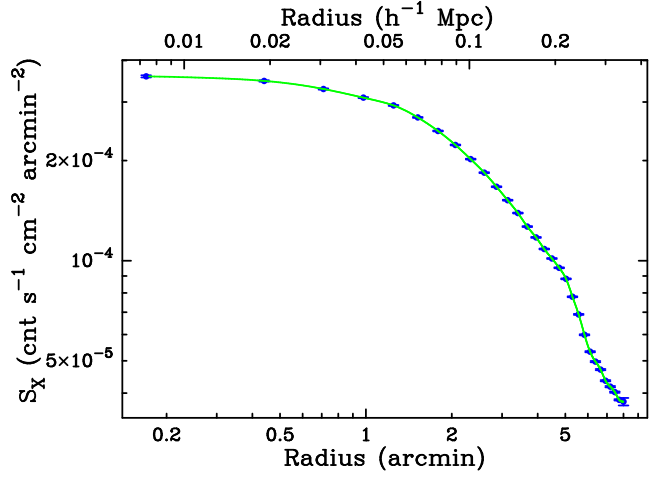}}
  \caption{\chandra radial profile of the galaxy cluster Abell
  3667 (points) with the interpolated model (solid line).  This model is
  used as the average X-ray surface brightness distribution in the
  calculation of $\dSx$.}
  \label{fig:a3667_radprof}
\end{figure}

The average X-ray surface brightness is required to compute $\dSx =
\Sx / \barSx$.  If one computes the average surface brightness,
$\barSx$, in annular shells, then one will tend to under (over)
estimate $\barSx$ toward the inner (outer) radius of each annulus.
Therefore, this will lead to an over (under) estimate of $\dSx$
toward the inner (outer) radius of each annulus.  To alleviate this
systematic, we adopt the azimuthally averaged X-ray surface brightness
as the model for $\barSx$, and use cubic spline interpolation between
radial bins.  The X-ray surface brightness radial profile for A3667 is
shown in Figure~\ref{fig:a3667_radprof}, along with the interpolated
model (line).

\begin{figure}[!tbh]
  \centerline{\includegraphics[height=6.3cm]{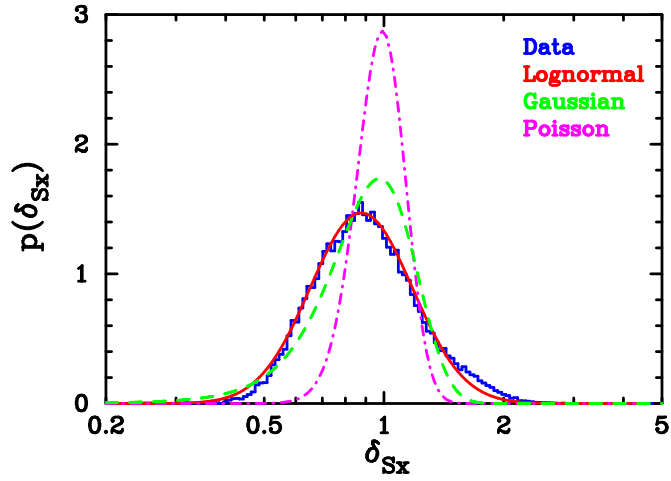}}
  \caption{Probability distribution of $\dSx$ from \chandra
  observations of the galaxy cluster Abell 3667 (blue histogram) along
  with the best fit lognormal distribution (red line) with $\SigSx =
  0.30$.  The lognormal distribution seems to be a reasonable
  description of the ICM inhomogeneity in A3667.  Also shown are the
  best-fit Gaussian model (dashed green) and a Poisson model
  (dot-dashed magenta) using the average counts per pixel within the
  fitting region.}
  \label{fig:a3667_hist}
\end{figure}

\begin{figure}[!tbh]
  \centerline{\includegraphics[height=6.3cm]{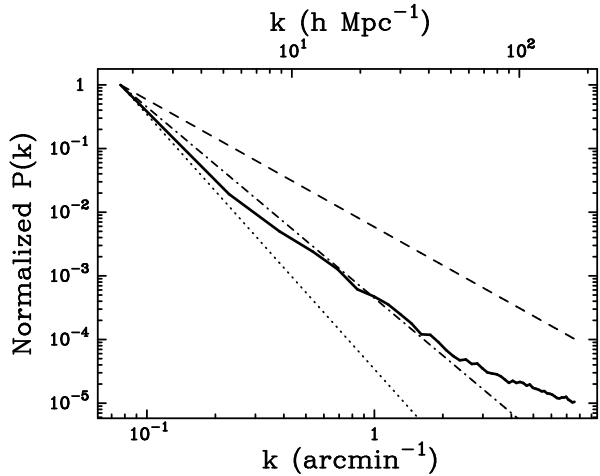}}
  \caption{Power spectrum of $\dSx$ (thick solid) from \chandra
  observations of the galaxy cluster Abell 3667, normalized to one at
  the largest scale.  Also plotted are three power-law power spectra
  with spectral indices of -2 (dashed), -3 (dot-dashed), and -4
  (dotted) for comparison.}
  \label{fig:a3667_ps}
\end{figure}

The probability distribution of $\dSx$, $p(\dSx)$, is computed from
the histogram of pixels calculated from the $\dSx$ image and shown in
Figure~\ref{fig:a3667_hist}.  The lognormal distribution
(Eq.~[\ref{eq:pdf_delta}]) is fit to the $p(\dSx)$ of A3667, where the
only free parameter is the standard deviation of the logarithm of
$\dSx$, $\SigSx$.  The best fit value for the lognormal model is
$\SigSx = 0.30$.  In addition, a Gaussian distribution is also fit to
the data, with its usual two parameters, the mean and standard
deviation.  Figure~\ref{fig:a3667_hist} shows the PDF of $\dSx$ for
the \chandra observations of the galaxy cluster A3667 (solid blue
histogram).  The best fit lognormal (solid red) and Gaussian (dashed
green) models are also shown.  A Poisson distribution (dot-dashed
magenta) is also shown for comparison, using the average counts per
pixel in the fitting region as the parameter for the Poisson
distribution.  Clearly, what is seen is not the result of Poisson
statistics.  The lognormal model seems to be a reasonable match to the
observed PDF.

However, without information on the power spectrum of the $\dSx$
fluctuations, it is difficult to interpret the value of $\SigSx$
(\S\ref{ss:sx_ps}) and relate it to the fluctuations in the density
distribution (Eqs.~[\ref{eq:fitsKa}, \ref{eq:Kalpha}];
Fig.~\ref{fig:mgsigout}).  Therefore, we take the Fourier transform of
the $\dSx$ image and compute the average power spectrum in wavenumber
annuli.  The power spectrum of $\dSx$ fluctuations is shown in
Figure~\ref{fig:a3667_ps} (thick solid) along with three power-law
spectra with spectral indices of -2 (dashed), -3 (dot-dashed), and -4
(dotted) for comparison.  The power spectrum of $\dSx$ has been
normalized to one at the largest scales.  A simple power-law model fit
to the power spectrum yields a spectral index of $\alphaSx = -2.7$
using the entire spectrum, and a spectral index of $\alphaSx = -3.0$
if excluding the larger wavenumbers ($\gtrsim 2$ arcmin$^{-1}$),
roughly where the power spectrum changes shape.

\subsection{Implications}
\label{subsec:obs_disc}

\begin{figure}[!tbh]
 \centerline{\includegraphics[width=95mm]{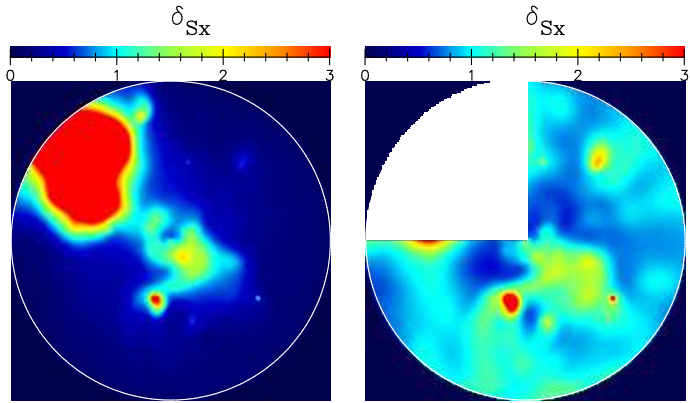}}
\caption{An example of a $\dSx$ map from a cosmological hydrodynamic
  simulated cluster (``Centaurus'') both before (left) and after
  (right) removal of a quadrant with a large clump.  Circles show the
  projected virial radius ($R_\mathrm{200}$).  Although within the
  projected virial radius, $R_\mathrm{200}$, these structures often
  reside outside of the three-dimensional virial radius,
  $r_\mathrm{200}$.}
\label{fig:clump}
\end{figure}

Both the standard deviation of the logarithm of X-ray surface
brightness fluctuations, $\SigSx = 0.30$, and the power spectrum
power-law index $\alphaSx \approx -3$, fall into the range expected
from hydrodynamical galaxy clusters and therefore used in the
synthesized cluster analysis (\S\ref{subsec:hydro_sim}).  By combining
these pieces of information, we can relate the information obtained
from the X-ray surface brightness distribution to that of the
underlying density distribution, using the results of the synthesized
cluster analysis.  Using the synthetic cluster result that the
spectral indices of the X-ray surface brightness fluctuations and that
of the Gaussian field are simply related by $\alphaSx \approx \alphaq
+0.2$, and the relation between $\Sign$, $\SigSx$, and $\alphaq$
(Eqs.~[\ref{eq:fitsKa}, \ref{eq:Kalpha}]; Fig.~\ref{fig:mgsigout}),
the \chandra results of $\SigSx = 0.30$ and $\alphaSx = -2.7$ imply
that the fluctuations in the underlying density distribution have
$\Sign = 0.43$.  A value of $\alphaSx = -3.0$ implies $\Sign = 0.36$.
The difficult test case of the A3667 X-ray surface brightness seems to
follow the lognormal distribution of density fluctuations, thus
enabling an estimate of the statistical properties of the underlying
ICM density fluctuations.

\section{Application to the Cosmological Hydrodynamic Simulated Clusters}
\label{sec:con}

Results from cosmological hydrodynamic simulations motivated the
lognormal model for ICM inhomogeneity.  In \S \ref{sec:synthetic}, we
found that synthetic clusters with lognormal fluctuations show a linear
relation between $\SigSx$ and $\Sign$.  We now return to clusters
extracted from cosmological hydrodynamic simulations to further explore
these results.

\begin{figure}[!tbh]
  \centerline{\includegraphics[width=120mm]{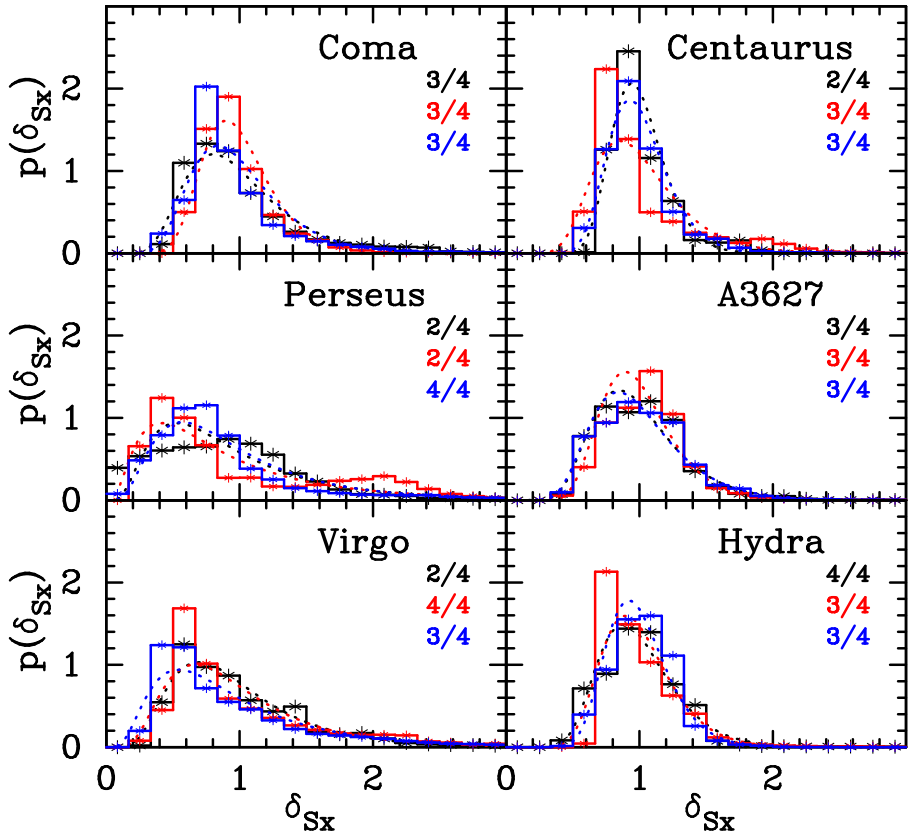}}
 \caption{The distribution of $\dSx$ for each of the six clusters from
  a cosmological hydrodynamic simulation (points and solid
  histogram). Each color indicates the projection along a different,
  orthogonal line of sight.  For each line of sight, we show the
  number of quadrants used for the analysis. For example, ``3/4''
  indicates that one quadrant is excluded and three remain.  The best
  fit lognormal model for each projection is also shown (dotted
  lines).
  \label{fig:chs}
}
\end{figure}

For each cluster extracted from the simulations, we create X-ray
surface brightness maps towards three orthogonal directions, and
compute $\dSx(\Rvc) = S_{\mathrm X}(\Rvc) / \barSx(R)$ in a similar
manner as described for the synthetic clusters in \S~\ref{ss:em}. The
regions we consider are within the projected virial radius $R_{\mathrm
200}$. The projected virial radius, $R_{\mathrm{200}}$, is the radius
within which the mean interior density is 200 times that of the
critical density.

\begin{figure}[!tbh]
 \centerline{\includegraphics[width=95mm]{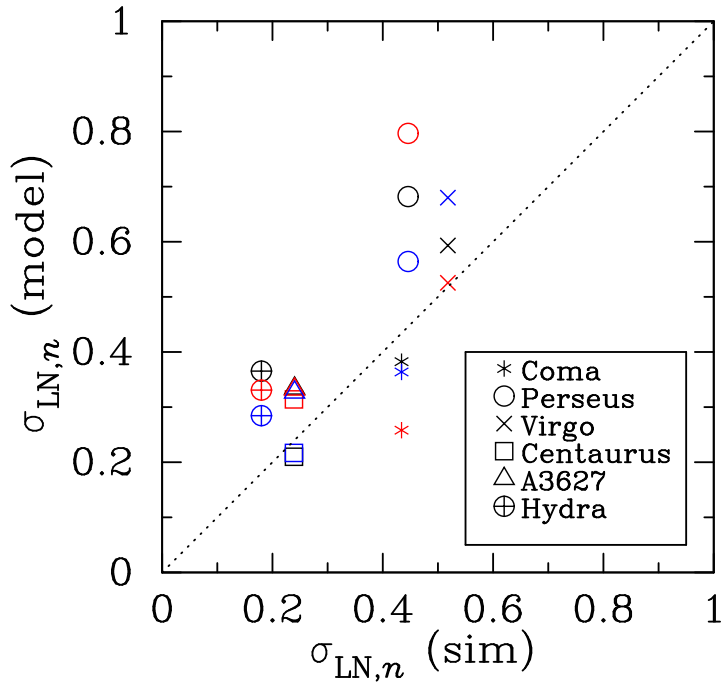}}
\caption{ The density fluctuation standard deviation predicted by our
  model, $\Sign(\mbox{model}) = \SigSx / Q (\alphaq)$ versus that from
  the simulations, $\Sign(\mbox{sim})$.  Symbols show different
  simulated clusters (see figure legend) and colors indicate different
  orthogonal lines of sight.  Also plotted is the simple linear
  relation $\Sign \mbox{(model)} = \Sign \mbox{(sim)}$ for comparison.
  }
  \label{fig:chsss}
\end{figure}

Although the lognormal distribution is a good fit to the density (and
temperature) of simulated galaxy clusters in three-dimensions, the
projection to X-ray surface brightness suffers from the additional
complexity of projection effects.  If large clumps are present, the
distribution of X-ray surface brightness fluctuations, $\dSx$, is not
well approximated by the lognormal distribution.  The large clumps
artificially distort the average profile of the cluster and therefore
bias the value of $\dSx$, which depends on the average profile.  We
also note that although these clumps fall within the projected virial
radius, $R_{\mathrm 200}$, they usually fall outside of the three
dimensional virial radius, $r_{\mathrm 200}$.  We therefore exclude
quadrants that contain large clumps, using $\dSx >10$ as the
exclusion criterion. Then, we recompute $\barSx (R)$ and $\dSx$. The
complex structure of simulated clusters is illustrated in the $\dSx$
images shown in Figure~\ref{fig:clump}, where examples of a simulated
cluster both before and after removal of a quadrant are displayed.
The circles show the projected virial radius, $R_\mathrm{200}$.

In Figure~\ref{fig:chs} the probability distributions of $\dSx$ for
the simulated clusters (histograms) along with the best-fit lognormal
model (dotted lines) are displayed. Each color indicates the
projection along a different, orthogonal line of sight.  Overall, the
probability distributions of $\dSx$ are reasonably well approximated
by the lognormal function, consistent with the results from the
synthetic clusters (\S~\ref{ss:synthetic_clusters}).

We now come full circle to compare our results from the synthetic
clusters directly to the simulations.  In order to do this, we look at
the relationship between $\Sign \mbox(\mathrm{sim})$ measured in the
simulated clusters and $\Sign \mbox(\mathrm{model})$ predicted from the
synthetic cluster results, equations~ (\ref{eq:fitsKa}) and
(\ref{eq:Kalpha}), where we adopt $\alphaq = \alphaSx - 0.2$ (see \S
\ref{ss:sx_ps}). The value of $\alphaSx$ for each simulated cluster is
obtained by fitting a power-law model, $P(K) \propto K^{\alphaSx}$, to
the power spectra of $\dSx$.  Because the resolution of the simulations
is much poorer than that of the synthetic clusters, we must recompute
the coefficients $c_1$ and $c_2$ in equation \ref{eq:Kalpha} from a set
of lower resolution synthetic clusters. Assuming $\rc \sim 100$ $h^{-1}$
kpc for the simulated clusters, we choose the resolution $\sim 0.1
\dgrid/\rc$, noting that this value corresponds to the {\it maximum}
resolution of the simulations.  Performing the same procedure described
in \S\ref{sec:synthetic}, we obtain $c_1 = 3.99 \times 10^{-2}$ and
$c_2= 3.36 \times 10^{-2}$.

We compare $\Sign \mbox{(model)}$ and $\Sign \mbox(\mathrm{sim})$ in
Figure~\ref{fig:chsss}.  Each color corresponds to a different line of
sight.  Although there is large scatter, these results indicate that
it is possible to estimate $\Sign$ within a factor of two only using
the information obtained from the X-ray surface brightness
distribution.

\section{Summary}
\label{sec:sum}

We have developed a method of extracting statistical information on the
ICM inhomogeneity from X-ray observations of galaxy clusters.  With a
lognormal model for the fluctuations motivated by cosmological
hydrodynamic simulations, we have created synthetic clusters, and have
found that their X-ray surface brightness fluctuations retain the
lognormal nature.  In addition, the result that $\SigSx$ and $\Sign$ are
linearly related implies that one can, in principle, estimate the
statistical properties of the three dimensional density inhomogeneity
($\Sign$) from X-ray observations of galaxy clusters ($\SigSx$ and
$\alphaSx$).

We have compared the predictions of our model to \chandra X-ray
observations of the galaxy cluster A3667.  For the first time in a
real galaxy cluster we were able to detect the lognormal signature of
X-ray surface brightness fluctuations, which was originally motivated
by simulations. Based on the synthetic cluster results, this enabled
an estimate of the statistical properties of the inhomogeneity of the
ICM of A3667.  In the context of lognormally distributed
inhomogeneity, we obtain $\Sign \approx 0.4$ for the gas density
fluctuations of A3667.  It is encouraging that the value of the
fluctuation amplitude for Abell 3667 is in reasonable agreement with
typical values from the simulated clusters.

Finally we check the validity and limitation of our method using
several clusters from cosmological hydrodynamic simulations. Unlike
the fairly idealized synthetic clusters, simulated clusters exhibit
complex structure more akin to real galaxy clusters.  As a result, the
empirical relation between the two- and three-dimensional fluctuation
properties calibrated with synthetic clusters when applied to
simulated clusters shows large scatter.  Nevertheless we are able to
reproduce the true value of the fluctuation amplitude of simulated
clusters within a factor of two from their two-dimensional X-ray
surface brightness alone.

Our current methodology combined with existing observational data is
useful in describing and inferring the statistical properties of the
three dimensional inhomogeneity in galaxy clusters.  The fluctuations
in the ICM have several implications in properly interpreting galaxy
cluster data. In particular, our current model may be useful in
interpreting data from current and future galaxy cluster surveys using
the Sunyaev-Zel'dovich effect, which have the potential to provide
tight constraints on cosmology.

\acknowledgments 

We thank Naomi Ota, Noriko Y. Yamasaki, and Kazuhisa Mitsuda for
useful discussions and Klaus Dolag for providing a set of simulated
clusters.  HK is supported by a JSPS (Japan Society for Promotion of
Science) Grant-in-Aid for science fellows. EDR gratefully acknowledges
support from a JSPS Postdoctoral Fellowship for Foreign Researchers
(P07030). This work is also supported by Grant-in-Aid for Scientific
research from JSPS and from the Japanese Ministry of Education,
Culture, Sports, Science and Technology (Nos. 20$\cdot$10466,
19$\cdot$07030, 16340053, 1874012, 20340041, and 20540235), and by the
JSPS Core-to-Core Program ``International Research Network for Dark
Energy''.

{\appendix

\section{Relation of Density and Surface Brightness Distributions
  Under the Thick-slice Approximation}
\label{a1:den_sb}

Modeling galaxy clusters with a spherical isothermal $\beta$ model
(Eq.~\ref{eq:ave1}), the surface brightness at an arbitrary projected
angular radius, $\theta$, is given by
\begin{eqnarray}
 \Sx(\theta) &\propto&  \int_{-\infty}^{\infty} [n (\rv)]^2  dl 
             \label{eq:a_sx1}\nonumber \\
             &=& \int_{-\infty}^{\infty}  \Dnn (\rv) \A n^2 \E
                   (r=\sqrt{l^2+\dAAtt})  dl
             \label{eq:a_sx2}\nonumber \\
             &=& M_2 n_0^2  \left( 1 + \frac{\dAAtt}{r_c^2} \right)^{-3 \beta}
	           \int_{-\infty}^{\infty} \Dnn (\rv) \left[ 1 +
	           \left( \frac{l}{\rceff} \right)^2 \right]^{-3 \beta} dl,
             \label{eq:a_sx3}
\label{eq:a_sx4}
\end{eqnarray}
where $\rceff \equiv \sqrt{r_c^2 + \dAAtt}$, and we assume the
$\sigma_{{\rm LX},x}$ in equation \ref{eq:pdf_delta} is independent of
$r$.  Therefore, the second moment of $n$ ($M_2 \equiv \A n^2 \E/\A n
\E^2 = \exp{(-\Sign^2)}$) is also independent of $r$. In the above, we
use $\A n \E  = \overline{n} (r)$.

The ensemble average of $\A \Sx(\theta) \E$ can be expressed as
\begin{eqnarray}
 \A \Sx(\theta) \E  &\propto&       \int_{-\infty}^{\infty} \A n^2 \E
                             (r=\sqrt{l^2+\dAAtt})  dl 
			     \label{eq:a_sx_avg2}\nonumber \\ 
                   &=&       \sqrt{\pi} n_0^2 M_2 r_c \frac{\Gamma(3
			     \beta - 1/2)}{\Gamma(3 \beta)} \left( 
                             1 + \frac{\dAAtt}{r_c^2} \right)^{-3 \beta + 1/2}
			     \label{eq:a_sx_avg3}
\end{eqnarray}

Combining equations (\ref{eq:a_sx3}) and (\ref{eq:a_sx_avg3}), $\dSxens$
reduces to
\begin{eqnarray}
\dSxens (\theta) &=& \kappa_\beta \int_{-\infty}^{\infty} \Dnn (\rv)
		     \left[ 1 + \left( \frac{l}{\rceff} \right)^2
		     \right]^{-3 \beta} d\left(\frac{l}{\rceff}\right)
		     \label{eq:a_dsx_ens} \nonumber  \\ 
\kappa_\beta     &\equiv& \pi^{-1/2}  \frac{\Gamma(3 \beta)}{\Gamma (3
                          \beta -1/2)}.
                          \label{eq:a_kappa_beta}
\end{eqnarray}

Now, fixing $\theta$, let us consider the three-dimensional field $\Dnn
(\rv) W_{\beta}(l)$ and its projected two-dimensional field $\dSxens
|_\theta$, defined as
\begin{eqnarray}
  \dSxens |_\theta&=& \int_{-\infty}^{\infty} \Dnn (\rv) W_{\beta}(\lprime)
  d\lprime  
  \label{eq:a_dsx_ens2} \\
W_\beta(\lprime) &\equiv& \kappa_\beta ( 1 + \lprime^2 )^{-3 \beta},
\label{eq:a_window}
\end{eqnarray}
where we use a dimensionless length normalized by $\rceff$ distinguished
by prime ($\lprime \equiv l/\rceff$, $\kzd \equiv \kz \rceff$).  Then,
we can consider the variance of the $\dSxens |_\theta$-field,
\begin{equation}
 \SNt = \frac{1}{(2 \pi)^2} \int d \Kvd P_{\Sxens |_\theta}(\Kvd),
\label{eq:s1}
\end{equation}
where $P_{\Sxens |_\theta}(\Kvd)$ is the (two-dimensional) power
spectrum of $\dSxens |_\theta$.  The variance of the $\Dnn$ field can
also be written as
\begin{equation}
 \Snt = \frac{1}{(2 \pi)^3} \int d \kvd P_{nn}(\kvd).
\label{eq:s2}
\end{equation}

With this, the relation between $P_{\Sxens}(\Kvd)$ and
$P_{nn}(\kvd)$ is
\begin{equation}
 P_{\Sxens |_\theta}(\Kvd) = \frac{1}{2 \pi} \int d \kzd
 P_{nn}(\kvd) |\widetilde{W}_\beta(\kzd)|^2.
 \label{eq:s3}
 \end{equation}
The Fourier conjugate $\widetilde{W}_\beta(\kzd)$ is given by
\begin{equation}
 \widetilde{W}_\beta(\kzd) = \kappa_\beta
 \left(\frac{2}{\kzd}\right)^{-3\beta+1/2} \frac{2 \sqrt{\pi}}{\Gamma
 (3 \beta)} K_{-3\beta+1/2} (\kzd),
\end{equation}
 where $K_{-3 \beta+1/2} (\kzd)$ is modified Bessel function of the
 second kind .

In the case that the largest scale fluctuation is smaller than the
physical scale (the thick-slice approximation, following \cite{FD04}),
the Fourier conjugate of the window function becomes the Dirac-delta
function, $|\widetilde{W}_\beta (\kzd)|^2 \sim g(\beta)
\delta(\kzd)$. The normalization factor $g(\beta)$ is given by
\begin{equation}
g(\beta) \equiv  2 \int_0^\infty d \kzd |\widetilde{W}_\beta (\kzd)|^2 = 2 \sqrt{\pi}
 \frac{\Gamma (3 \beta)^2 \Gamma (6 \beta -1/2)}{\Gamma(3 \beta
 -1/2)^2 \Gamma(6 \beta)}.
\end{equation}
 Let us define the effective width 
\begin{equation}
 \Deff (\theta) \equiv 2 \pi \rceff/ g(\beta) = \sqrt{\pi} \frac{\Gamma(3 \beta -1/2)^2 \Gamma(6 \beta)}{\Gamma (3 \beta)^2
  \Gamma (6 \beta -1/2)} \rceff.
\end{equation}
\cite{FD04} explore the column density distribution assuming a plane
parallel geometry with width $\Delta$. In the thick slice case,
$\Deff$ corresponds to $\Delta$ although they consider the column
density not the surface brightness.  We assume statistical isotropy
and a power law spectrum with upper and lower limit ($\kmaxd \equiv
\kmax \rceff$ and $\kmind \equiv \kmin \rceff$),
\begin{eqnarray}
P_{nn}(\kvd) 
      \left\{
       \begin{array}{lr}
           \propto |\kvd|^{\alphaDnn} &  \mbox{$\kmind<|\kvd|<\kmaxd$} \\
           = 0 & \mbox{otherwise.} \\
       \end{array}
      \right.
\end{eqnarray}

Finally, using equation (\ref{eq:s1}), (\ref{eq:s2}), and (\ref{eq:s3})
under the thick-slice approximation, we obtain
\begin{eqnarray}
\label{eq:thickr}
 \SNt/\Snt = 
      \left\{
       \begin{array}{lr}
\displaystyle{ \frac{1(\alphaDnn+3)(1-\zeta^{\alphaDnn+2})}{2(\alphaDnn+2)(1-\zeta^{\alphaDnn+3})}
  \left(\frac{\Deff (\theta)}{\lmax}\right)^{-1}} & \mbox{$\alphaDnn\neq -3$ and $\alphaDnn\neq -2$}\\
\displaystyle{ \frac{\log \zeta}{2 (\zeta -1)} \left(\frac{\Deff (\theta)}{\lmax}\right)^{-1} } & \mbox{$\alphaDnn = -2$} \\
\displaystyle{ \frac{1-1/\zeta}{2 \log \zeta} \left(\frac{\Deff (\theta)}{\lmax}\right)^{-1}} & \mbox{$\alphaDnn = -3$,} \\
       \end{array}
      \right.
\end{eqnarray}
where $\zeta \equiv \kmax/\kmin$ and $\lmax \equiv 2 \pi \kmin^{-1}$.

Then, although $\SNt$ is the variance of $\dSxens |_\theta$-field, we
regard it as the variance of the ensemble average of $\dSxens(\Rvcd)$ at
$\theta$. The conversion to the standard deviation of logarithm is
expressed as
\begin{eqnarray}
  \SigSx = \sqrt{\log{(1+\SNt)}}
\end{eqnarray}
In Figure~\ref{fig:rdt}, we adopt $\zeta= \Lbox/(2
\dgrid)=f_{\mathrm{s}}/f_{\mathrm{Ny}}$, where $f_{\mathrm{s}}$ and
$f_{\mathrm{Ny}}$ are the sampling frequency and the Nyquist frequency,
respectively, and $\lmax=\Lbox$. We also adopt the fitted value of
$\alphaDnn$ in equation (\ref{eq:thickr}).

\section{Distribution of the Density Squared}
\label{sec:a2_densquared}

If one assumes density inhomogeneity fluctuations, $\Dn = n/\A n \E$,
follow the lognormal distribution, $p_{\mathrm{LN}}(\Dn;\Sign)$, the
fluctuations of the density squared, $\Dnn \equiv n^2/\A n^2 \E$ can be
written as
\begin{eqnarray}
\Dnn = \Dn^2 \frac{\A n \E^2}{\A n^2 \E}  =  \Dn^2 \exp{(-\Sign^2)}.
\end{eqnarray}
where $\A \, \E$ indicates ensemble average.

For simplicity, we replace $\Dn$ and $\Dnn$ by $x$ and $y$, respectively,
\begin{eqnarray}
 x \equiv \Dn ;  y \equiv \Dnn.
\end{eqnarray}
Therefore, the relation between $x$ and $y$ is
\begin{eqnarray}
 x = \sqrt{y} \exp{(\Sign^2/2)}.
\end{eqnarray}
Because $x$ follows $p_{\mathrm{LN}}(x;\Sign)$, the distribution of
$y$ is obtained by
\begin{eqnarray}
 p(y) = p_{\mathrm{LN}}(x;\Sign) \frac{d x}{d y} =  p_{\mathrm{LN}}(y;2 \Sign).
\end{eqnarray}
Therefore, $\Dnn$ follows the lognormal distribution with lognormal
standard deviation of $2 \Sign$.
}


\end{document}